\documentclass{elsart}
\usepackage{mathrsfs}
\usepackage{feynmf} 
\def\ie{\hbox{\it i.e.}{}}
\def\eg{\hbox{\it e.g.}{}}
\def\etc{\hbox{\it etc}{}}
\def\nn{\hspace{2mm}}
\def\vev#1{\left\langle #1\right\rangle}
\def\Tr{{\rm Tr}{}}
\begin{document}
\runauthor{H.~B.~Nielsen, Y.~Takanishi}
\hfill
\vbox{
\halign{#\hfil\cr
IC/2003/116    \cr
hep-ph/0310146 \cr
\cr }}
\begin{frontmatter}
\title{Green-Schwarz Anomaly Cancellation, World Sheet Instantons 
and Wormholes}
\author[NBI]{H.~B.~Nielsen},
\ead{hbech@alf.nbi.dk} 
\author[ICTP]{Y.~Takanishi}
\ead{yasutaka@ictp.trieste.it}
\address[NBI]{The Niels Bohr Institute, Blegdamsvej 17, 
DK-2100 Copenhagen {\O}, Denmark}
\address[ICTP]{The Abdus Salam International Centre for Theoretical
Physics,\\ Strada Costiera 11, I-34100 Trieste, Italy}
\begin{abstract}
  We consider the breaking of the global conservation of gauge field
  charges which are commonly thought to survive the spontaneous
  breakdown of gauge symmetry brought about by Kalb-Ramond fields.
  Depending on the dilaton field and also the size of the
  compactifying space, the global charge breaking may take place due
  to world sheet instantons. In going to $3+1$ dimensions one could
  have a serious problem in order to produce the hierarchies between
  the quark and the charged lepton masses using the mass protecting
  charges with the Green-Schwartz anomaly cancellation. Various
  unnatural features of this type of models are discussed.
\end{abstract}
\end{frontmatter}

\section{Introduction}
\label{sec:intro}

It has been believed some time ago that gauge symmetries with
anomalies cancelled by the Green-Schwarz mechanism~\cite{GS} in $9+1$
dimensions could be spontaneously broken in such a way that, after
compactifying to four dimensions, a global symmetry inherited from the
gauge symmetry could, in principle, survive,
$\eg$,~\cite{GS1,GS2,GS3,GS4,GS5,globalgs1,globalgs2,globalgs3,globalgs4,globalgs5,globalgs6,globalgs7,Petcov,globalgs8,Lola,globalgs9,globalgs10,globalgs11,globalgs12,globalgs13,globalgs14,globalgs15,globalgs16,globalgs17,globalgs18,globalgs19,globalgs20,globalgs21}. %

The survival of a global gauge symmetry -- even after the gauge bosons
have acquired a Higgs field induced mass -- is quite mysterious
because the vacuum is not gauge invariant under the gauge symmetry
with constant gauge function ($\Lambda=\mathrm{const.}$) due to the
Kalb-Ramond field (which plays a major role in this phenomenon).
Despite this, it seemed as though there remains a phase
transformation symmetry for the fields carrying the family-dependent
$U(1)_X$-charge when gauge symmetry is spontaneously broken in this
remarkable way.

We emphasise in this article that, precisely this phase
transformation symmetry is {\em not} a true symmetry if one takes into
account the world sheet instantons. We shall argue below that if the
compactifying space is of order of the fundamental scale, then the
local and global gauge symmetry gets totally broken. However, it is
unrealistic to compactify the space so close to the fundamental scale.
The crucial point is that the effect of the world sheet instantons will be
exponentially suppressed. In the case of very strong breaking (of the
order of the fundamental scale) it would mean that we could not apply
the Green-Schwarz anomaly cancellation mechanism in $3+1$ dimensions,
$\ie$, for the application to the large hierarchical Yukawa coupling
constant structures.

On the other hand, if we let the compactifying scale be much below the
fundamental scale, there is another odd feature: there is very light
(Abelian) gauge particle from the fundamental scale point of view and
correspondingly we get that the condition 
$F_{\mu\nu}\widetilde{F}^{\mu \nu}=0$ for not having anomalies is fulfilled
identically. This would lead to a rather strange electrodynamics. If
the validity of the usual $\mathrm{div} \vec{E}=j^0$ is not
maintained, the dynamics could open the possibility for space-time
foam causing the break down of the global charge conservation due to
wormholes.

This article is organised as follows: in the next section, we review
the Green-Schwarz anomaly mechanism, and in section $3$ the world
sheet instantons, and we also discuss the string coupling constant. Section 
$4$ contains the various discussions including the suspected
effects of wormholes. Finally, section $5$ contains our conclusions.

\section{Review of Green-Schwarz anomaly mechanism}
\label{sec:greenschwarz}
 
Let us review the Green-Schwarz anomaly cancellation mechanism
focusing on the Kalb-Ramond field in $9+1$ dimensions and then
the application for the $3+1$ dimensions.

{}For the purpose of making phenomenological fits of the quark and
(charged) lepton masses and mixing angles it is very useful to have
some approximately conserved charges~\cite{FN} (in addition to the
gauge charges of the Standard Model) so that most of the masses get
suppressed due to the differences in quantum numbers of the right- and
left-handed Weyl components. It is very attractive, and needed due to the
effects of wormholes~$\etc.$~\cite{Gilbert}, to let such mass
suppressing charges be gauge charges. There are many gauge charges in 
superstring theory so such a picture is not unnatural 
in this theory. Working in $3+1$ dimensions one
would at some level expect to obtain a $3+1$ dimensional field
theory with gauge fields which could be described as renormalisable.
That in turn would imply that the triangle anomalies resulting from
the various chiral fermions in the effective $3+1$ dimensional model
should cancel, $\ie$, no violation of gauge symmetry would be caused.
Otherwise this effective model would not be renormalisable. Now,
however, it became very popular to use the inspiration from the
superstring theory to suggest models in which this ``usual'' gauge
anomaly cancellation does {\em not} take place. From the four
dimensional point of view this avoidance, which is
usually needed for renormalisation, gauge- and mixed anomaly cancellation
conditions seems quite extraordinary: A certain coefficient 
field $b(x^\mu)$ 
in an expansion for the Kalb-Ramond field $B_{MN}~(M,N=0,1,\ldots, 9)$ in
the $9+1$ dimensional theory, couples as an axion field. That
is to say it couples via the Lagrangian density term of the form
\begin{equation}
  \label{eq:axionlag}
  \mathscr{L} = b(x^\mu)\, F_{MN}\, \widetilde{F}^{MN}\, + \cdots \nn,
\end{equation}
where $\mu=0,1,2,3$.

In the superstring theories (type II and Heterotic strings) there
is a Kalb-Ramond anti-symmetric tensor field with two indices 
on the potential $B=B_{MN}\, d x^M\wedge d x^N$ and three on its 
field~\cite{GS}
\begin{equation}
  \label{eq:Hgs}
  H = d B + \omega_{3 Y}^0 - \omega_{3 L}^0 \nn.
\end{equation}
Here the three forms $\omega_{3 Y}^0$ and $\omega_{3 L}^0$ 
are given by~\cite{Zumino1,Zumino2}.

In these theories there is a very sophisticated
way of cancelling the gauge, gravitational and various mixed anomalies,
first by having the right number of chiral fermions but in addition
some terms, which are gauge non-invariant when alone, in the action for zero
mass particles are used,
\begin{eqnarray}
  \label{eq:gsaction1}
  &S_1 =& \phantom{-} c \int \left( \, B \, \mathrm{tr}{} F^4 
+ \frac{2}{3}\, \omega_{3 Y}^0 \, 
\omega_{7 Y}^0 \right) \nn, \\
\label{eq:gsaction2}
  &S_2 = & - c \int \left[ \frac{1}{32} \, B \, \left( \mathrm{tr}{}\ R^2 
\right)^2 
+ \frac{1}{8} \, B\, \mathrm{tr}{}\ R^4 + \frac{1}{12} \,  
\omega_{3 L}^0 \, \omega_{7 L}^0 \right] \nn, \\
\label{eq:gsaction3}
 &S_3 = & \phantom{-} c \int \left( \frac{1}{8} \, B \, \mathrm{tr}{}\ R^2 \, 
\mathrm{tr}{}\ F^2 + \frac{1}{48}\, \omega_{3 L}^0 \, \omega_{3 Y}^0 \,
 \mathrm{tr}{}\ R^2 
- \frac{1}{24} \, \omega_{3 Y}^0 \, \omega_{3 L}^0 \, \mathrm{tr}{}\ F^2 
\right. \nonumber \\ 
&& \phantom{S_3 = } - \frac{2}{3}\, \omega_{3 L}^0\, \omega_{7 Y}^0 
+ \frac{1}{12}\, \omega_{3 Y}^0 \, \omega_{7 L}^0 \left. 
\phantom{\mathrm{tr}{} R^5 F^2} \hspace{-1.4cm} \right) \nn,
\end{eqnarray}
to cancel the remaining part of the anomalies. Here $c$ is numerical
constant~\cite{Gano,Zumino1}.

In order to get chiral fermions -- as is phenomenologically required
to obtain the Standard Model -- it is needed to break the parity
symmetries that only makes reflections in the compactifying
dimensions (for instance by having non-zero magnetic field in the
extra dimensions). One may typically make use of Calabi-Yau spaces as
the $6$-dimensional compactifying space. For pedagogical reasons, just
to illustrate the idea we may in the present article think of a
compactifying space being the cross product of three spheres, each of
topology $S^2$ and each with a magnetic field on them corresponding to
a ``magnetic monopole in the centre of the $S^2$ sphere''. Let us
imagine that the equations of motions have led to that the vacuum has
$S^2$ rotation invariant fields on the different $S^2$'s. Then we may
symbolically use these rotational
invariant%
\footnote{%
  This $S^2$ rotation symmetry is, we have in mind, that the
  topological $S^2$ is represented by a sphere which then
  has the symmetry under $SO(3)$ rotations (about a point outside the
  sphere) with respect to the fields assumed to be present.}
field strength 
$\vev{F_{67}}$~$\etc$. The first term in integrand of 
$S_1$ (Eq.~\ref{eq:gsaction1}) which by itself gauge breaking, will
contain a contribution of the form
\begin{equation}
  \label{eq:axtionlow}
   \mathscr{L} = B_{45}\, \vev{F_{67}}\,\vev{F_{89}}
\, F_{\mu\nu}\, \widetilde{F}^{\mu\nu} \, + \cdots\nn. 
\end{equation}
Imagine expending the variation over the $S^2$ sphere ($\ie$, $x^4$
and $x^5$ dependence) on ``spherical harmonics'' or ``eigenfunctions''.
Suppose we arranged one of spherical harmonic or eigenfunction to be
dominant in the weakly exited state. We describe the effective four
dimensional theory by means of the expansion coefficient $b(x^\mu)$ to
this term:
\begin{equation}
\label{eq:expb45}
  B_{45}(x^\mu, x^4, x^5) = f(x^4, x^5) \, b(x^\mu) \nn.
\end{equation}
Really we could define such $b(x^\mu)$ by integrating the two form $B$
over a homotopically non-trivial $2$-cycle. This would then require
that we imposed other terms in the expansion of $B_{45}$ 
to be restricted to zero. Taking the
magnetic fields in the compactifying dimensions as constants we end up
with an effective term in the four dimensional Lagrangian density
which up to the over all constant is of the form 
(\ref{eq:axionlag}). From the kinetic term for the Kalb-Ramond field,
\begin{equation}
  \label{eq:kinKR}
  -\frac{3 \kappa^2}{2 g^4 \varphi^2}\, H_{M N P} \, H^{M N P} \nn,
\end{equation}
where $\kappa$ is the gravitational coupling constant, 
$\varphi$ the dilaton field, $g$ the Yang-Mills gauge coupling constant
in the Lagrangian density, we obtain a kinetic term for the 
coefficient field $b(x^\mu)$. Due to the 
$\omega_{3 Y}^0$ term in Eq.~(\ref{eq:Hgs})
it comes together with an Abelian part of the Yang-Mills potential in
an expression of the form
\begin{equation}
  \label{eq:YMKin}
  \frac{1}{2} \, m^2 \, \left( \partial_\mu b - A_{\mu} 
\right)^2 \nn.
\end{equation}
This is a gauge invariant combination for
the $b$-field gauge transform
\begin{equation}
  b \rightarrow b + \Lambda \nn,
\end{equation}
while 
\begin{equation}
  A_\mu \rightarrow A_\mu + \partial_\mu \Lambda \nn,
\end{equation}
where 
$\Lambda$ is the gauge function for an invariant $U(1)$ subgroup 
of the left over symmetry group, not spontaneously broken.

{}For simplicity we imagine that the presence of the extra dimension
fields $\vev{F_{67}}$ $\vev{F_{89}}$ represents a break down to a
subgroup which still contains at least one invariant Abelian subgroup
called $U(1)_X$. Then we may concentrate on the gauge field associated
with this subgroup $U(1)_X$ and denote the gauge function for it as
$\Lambda$. We shall discuss a rather extraordinary behaviour of the
theory -- from the four dimensional point of view -- with the axion
field $b(x^\mu)$ (see also Sec.~\ref{sec:foureffec}).

Suppose that $b(x^\mu)$ does not quantum fluctuate so widely that it
totally looses an expectation value. This means that there is a
spontaneous break down of the gauge symmetry for $U(1)_X$. Since even
the constant $\Lambda$ gauge transformation is spontaneously broken
due to the additive transformation property of $b$, the spontaneous
breaking situation is just like that of the Higgs case. However, that
means, one would expect that particles -- such as fermions carrying
$U(1)_X$-charge quanta -- would be able to make transitions into (sets
of) particles with a different number of such charges (together). At
this point, however, one has often found the belief that the global
symmetry and the Noether conservation of the charged particles is {\em
  not} violated. In the perturbative approximation this belief is
well-granded. We will discuss this question in the following in 
non-perturbative approximation.

\section{World sheet instantons and the Fayet-Iliopoulos D-term}
\label{sec:worldsheetinst}

Although at first it seems as if there is no way to cause the 
global $U(1)_X$-charges on particles to be created or
annihilated, it was shown in~\cite{worldsheetinst1,worldsheetinst2}
that such a violation of the charge was indeed occurring due to world
sheet instantons. These ``world sheet instantons'' refer to the
tunnelling of a string so as to have a ``time track'' during
tunnelling which encloses in our simple scenario the $S^2$ involved
with the $B_{45}$. In the real general case we should have the
tunnelling go around a $2$-cycle homotopical to the $2$-cycle(s) used
for extracting $b$ from $B$. 

The important point for the present discussion is as follows: When such a
world sheet instanton exists, there are zero modes for the
fermions (as well as bosons) which have $U(1)_X$-charge. These zero
modes cause the $U(1)_X$-charge to change. In this anomalous way -- 
similar to the QCD-instanton -- the global charge gets also
violated after all. According 
to~\cite{worldsheetinst1,worldsheetinst2} this is much more natural  
than not having the global $U(1)_X$-charges broken.

In reality the effect of this zero-mode effect is described by an
effective Lagrangian term
\begin{equation}
  \label{eq:instop}
  e^{-i b}\, Q = e^{-i b} Q_1 \cdot Q_2 \cdots Q_n \nn.
\end{equation}
Here the $Q_1, Q_2, \dots, Q_n$ are various $U(1)_X$-charged
fields and the product $Q=Q_1 \cdot Q_2 \cdots Q_n$ could, for
instance, be $Q={\overline\psi} \psi$ where $\psi$ is a field for which 
$U(1)_X$ has the role of a chiral mass protecting charge. The 
whole term is to be multiplied
by the amplitude of the world sheet instanton. The factor
$e^{-i b}$ comes from the exponentiated string action 
Eq.~(\ref{eq:insttunn}), which may cause the 
damping and the factors $Q=Q_1\cdot Q_2 \cdots Q_n$ 
are needed to make the whole term gauge invariant.

In supersymmetric theories, where these models are usually
considered, such an axion field must occur together with a
corresponding field in a complex combination. For instance the
anomalous Fayet-Iliopoulos D-term was studied~\cite{GS3} in the
context of the heterotic string theory: They consider a dilaton chiral
supermultiplet, $\Phi$, adding to the K{\"a}hler potential a part,
$K_d\left(\Phi+\overline{\Phi}\right)$, which transforms under the $U(1)_X$
gauge group as
\begin{equation}
  \label{eq:U(1)xtra}
  \delta \Phi = i \alpha \nn.
\end{equation}
The K{\"a}hler potential is with appropriate vector kinetic term 
\begin{equation}
  K= -\ln(\Phi + \overline{\Phi})\,, \qquad  f= \Phi \,, 
\qquad \Phi \equiv \phi^{-2}  + i  b\,.
 \label{eq:stringyK}
\end{equation}
The gauge coupling constant thus depends on the dilaton, 
$({\rm Re}f)^{-1}=\phi^2$. The theory also has an axion coupling
proportional to $b\,F_{\mu\nu}\,{\widetilde F}^{\mu\nu}$. With the
shift transformation of the axion field under $U(1)_X$ gauge group
(see Eq.(\ref{eq:U(1)xtra})), this term serves to remove the anomaly
proportional to $F_{\mu\nu}\, {\widetilde F}^{\mu\nu}$.

The D-term potential is using Eq.~(\ref{eq:stringyK})
\begin{equation}
\frac{1}{2} ({\rm Re} f)^{-1} 
\left\{ K_d^\prime\left(\Phi + \overline{\Phi}\right) \right\}^2 
= \frac{1}{8} \phi^6 \nn.
\label{eq:heterotic}
\end{equation}
We should emphasise that in the weak coupling limit, where dilaton
field goes to zero ($\phi^2\to0$) the Fayet-Iliopoulos D-term
vanishes, $\ie$, there is a supersymmetric minimum.

If we contrary to the just given argumentation assumed that after all
non-zero Fayet-Iliopoulos D-term were stabilised -- $\ie$, there were
at least the metastable minimum for non-zero $\phi$ -- then for the
weakly-coupled heterotic string according to Ref.~\cite{GS4} with
$g_s=\phi^2$, there is a Fayet-Iliopoulos D-term given by
\begin{equation}
\xi_{GS}  = \frac{g_s^2 \, \Tr q}{192 \pi^2}  M^2_{Pl}\nn,
\label{eq:anomalousFI}
\end{equation}
where $M_{Pl}$ is the Planck mass and $\Tr q$ is the sum of $U(1)_X$-charges.

Thus the potential energy becomes
\begin{equation}
  \label{eq:potential}
  V_D= \frac{1}{2}\, g_s^2 \xi_{GS}^2 
= \frac{\left(\Tr q\right)^2}{73728 \pi^4} \, g_s^6 \, M_{Pl}^4 \nn, 
\end{equation}
so that $V_D$ is obviously negligible at very small dilaton field even
if one wishes to have a model in which $\Tr q$ is non-zero, a typical
value of $\Tr q\sim 10^2$ to $10^3$ (see $\eg$~\cite{globalgs3}).

According to~\cite{globalgs6} the D-term~(\ref{eq:anomalousFI}) may
induce some (non-zero) expectation value of a scalar field called
$\theta$, and thus cause a spontaneous breakdown of the global (part)
of the $U(1)_X$ symmetry. Supposing no other scaler field break this
group, then the non-zero vacuum expectation value of $\theta$ may be
made a connection with the Fayet-Iliopoulos D-term in following way:
\begin{equation}
\label{eq:theta}
\left|\!\vev{\theta}\!\right| 
= \sqrt{-\frac{\xi_{GS}}{X_\theta}} =\sqrt{\xi_{GS}}\nn,
\end{equation}
where the $U(1)_X$-charge of the Higgs field $\theta$, $X_\theta$, is
taken to be $-1$. In this way one could obtain an expansion parameter
for the fermion mass matrices from $\xi_{GS}$: In the case of
$\xi_{GS}$ being not too large (see Eq.~(\ref{eq:anomalousFI})), in
other words, $g_s$ is not too small and the sum of $U(1)_X$-charges ($\Tr q$)
is of order $10^2$ to $10^3$, we could use such a mechanism to produce
a good order of magnitude for the breaking of the global charge
conservation for $U(1)_X$. The $U(1)_X$ symmetry would be broken one
or two orders of magnitude below the fundamental scale (Planck or
string scale, depending on model). This situation could be using 
the fundamental scale as the Planck scale, 
\begin{eqnarray}
\label{eq:epsilon}
\epsilon&\equiv& \frac{\left|\!\vev{\theta}\!\right|}{M_{Pl}} \nonumber \\
&=&\sqrt{{\xi_{GS}\over M^2_{Pl}}} \sim 0.2 \nn,
\end{eqnarray}
$\ie$, an expansion parameter for the fermion mass matrices which is
identified with the Cabibbo angle.

If we have identified the string coupling constant, $g_s$, with a
dynamical field -- the dilaton field $\phi^2$ -- the ground state will
be found by adjusting this coupling $g_s$ to be zero, which obviously
means the disappearance of the global charge breaking effect. In this
situation, which is the expected one, we have thus at first no
complain from world sheet instantons about the possibility hoped for
in literature of having the global charge totally conserved although
derived from a Higgs gauge symmetry. However, if the string coupling
constant $g_s$ is really going to zero, then the models in which this
happens have zero string coupling constant, and they become 
{\em totally free}, since in string theory all interactions 
are finally
derived from the string interaction $g_s$. Unless one can somehow 
live with a tiny breaking of supersymmetry at very high energy
scale allowing a small but non-zero $g_s$, the models avoiding a
Fayet-Iliopoulos D-term would become totally free as string
theories!

The wish that the adjustment lets the Fayet-Iliopoulos D-term vanish
by adjusting the $\phi^2$ to be zero is further supported by the often
favored phenomenological call for supersymmetry not to be broken
except at low energy scales, typically close to 1 TeV, because it can
be of help for the hierarchy problem. From the point of view of the
high scale of energy, where we \mbox{\it a priori} have the Higgsing
of the $U(1)_X$ group, a phenomenologically useful supersymmetry
breaking scale would be tremendously small and essentially count as
zero. This argument further disfavours models which would have
troubles due to the Fayet-Iliopoulos D-term. But if the problem is
solved by adjusting to the supersymmetry which is only extremely
weakly broken by having the string coupling almost zero, then the type
of model is even more problematic, because it lacks the interaction
altogether.

As a resume of the above discussion let us compare the two logical
possibilities concerning the size of the string coupling constant
$g_s$ of being of order unity $g_s\approx \mathscr{O}(1)$ or sufficiently small
($g_s\approx0$) both being considered in the presence of a
non-trivial Green-Schwarz anomaly cancellation so that the
Fayet-Iliopoulos D-term becomes non-zero unless $g_s=0$:
\begin{enumerate}
\item[{\it (1)}] Consistence 
\item[] The possibility $g_s\approx \mathscr{O}(1)$ is strictly speaking
  inconsistent because the Fayet-Iliopoulos D-term
  (Eq.~(\ref{eq:anomalousFI})) drives the $g_s$, which is effectively
  dynamical (related to $\phi$), to zero so that only $g_s\approx0$ is
  consistent. 
\item[{\it (2)}] The expansion parameter $\epsilon$ suitable for
  small hierarchy?
\item[] A reasonable sized $g_s\sim 1$ could give a good expansion
  parameter $\epsilon$, however, if $g_s\approx 0$ of course
  $\epsilon$ becomes exceedingly small and not useful for fitting the
  small hierarchy. 
\item[{\it (3)}] Supersymmetry for hierarchy problem phenomenology?
\item[] $g_s\approx \mathscr{O}(1)$ leads to extremely high scale supersymmetry
  breaking by the Fayet-Iliopoulos D-term such that the supersymmetry is of
  no use for solving the hierarchy problem. (explain any weak scale
  supersymmetry phenomenology for that matter.) While for a very small
  $g_s\approx 0$ one may get supersymmetry accurate enough for
  hierarchy problem purposes.
\item[{\it (4)}] Freeness of the whole string theory
\item[] For $g_s\approx 0$ the string theory becomes free while for
  $g_s\approx \mathscr{O}(1)$ there are interactions.
\item[{\it (5)}] Conservation of the global charges of $U(1)_X$
\item[] There are two possibilities with $g_s$ in the relative large
  range which we should mention:
\begin{enumerate}
\item[{\it (a)}] $g_s$ is large: in this case the world sheet
  instantons and the Fayet-Iliopoulos D-term may be too large so that
  the breaking, which they cause, would also be too large ($\ie$ the
  expansion parameter $\epsilon$). In this case $U(1)_X$-charge cannot
  be used to the small hierarchy problem of the fermion masses.
\item[{\it (b)}] $g_s$ is not too large: in this case $g_s$ could give
  a good expansion parameter $\epsilon$ (Eq.~(\ref{eq:epsilon}) and
  see Sec.~\ref{sec:order}). Then one could hope for the desired
  breaking of global charge {\em without} invoking further breaking
  mechanisms.
\end{enumerate}
In the case when $g_s$ is really small the 
global charge is very well conserved, but one can imagine it broken by 
other means so that it would be no problem.
\item[{\it (6)}] Higgsing
  
\item[] The Higgsing of the local part of the gauge group uses the
  Kalb-Ramond field and that works independent of the size of $g_s$
  (unless of course the whole theory becomes free that one can
  conclude that nothing happens at all).
\item[{\it (7)}] Anomaly cancellation for $U(1)_X$ from four
  dimensional point of view
\begin{enumerate}
\item[{\it (a)}] a strong $g_s$: In this case the world sheet
  instanton breaking would be strong and there would essentially be
  not even a global charge approximately conserved.
\item[{\it (b)}] a bit weaker $g_s$: there is an approximately
  conserved global charge which does not have the cancellation by
  triangle diagrams which is required for a gauge charge.
\end{enumerate}
\end{enumerate}

\section{Discussions}
\label{sec:discussion}
\subsection{The extraordinary properties of the 
four dimensional effective theory}
\label{sec:foureffec}

The four dimensional model is derived from an although
non-renor\-malis\-able -- as all theories in ten dimensions -- then at
least gauge invariant theory. It is therefore surprising that it does
not satisfy the usual conditions on numbers of fermion species and
their charges needed for the anomaly cancellation. It is the reason why 
there is the Wess-Zumino term in Eq.~(\ref{eq:Hgs}). However, how does
that remove the need for anomaly cancellations?

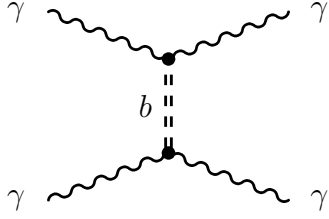
\begin{figure}[t!]
\unitlength=1mm  
\begin{center}
\vspace{3mm}
\begin{fmffile}{Photonb}
\begin{fmfgraph*}(40,25)
\fmfleft{i1,i2}
\fmfright{o1,o2}
\fmfv{label=$\gamma$,l.a=180,l.d=3mm}{i1}
\fmfv{label=$\gamma$,l.a=180,l.d=3mm}{i2}
\fmfv{label=$\gamma$,l.a=0,l.d=3mm}{o1}
\fmfv{label=$\gamma$,l.a=0,l.d=3mm}{o2}
\fmf{photon}{i1,v1,o1}
\fmf{photon}{i2,v2,o2}
\fmf{dbl_dashes,label=$b$}{v1,v2}
\fmfdot{v1,v2}
\end{fmfgraph*}
\end{fmffile}
\vspace{5mm}
\end{center}
\caption{\protect Feynman diagram of the photon-photon scattering by 
$b$ field exchange.}
\vspace{5mm}
 \label{fig:photonb}
\end{figure}

One might wonder how the situation of the anomaly cancellation would
be, if we had that the mass scale $m$ of the $U(1)_X$-photon
(Eq.~(\ref{eq:YMKin})) were very low compared to the scale of energy at
which we consider the situation. From the point of view of 
such a high energy scale compared to $m$, it would
seem that the kinetic term for the axion field, $b$, were very close to
having zero coefficient, $\ie$, an auxiliary field. The effect of
integrating out $b$ functionally would be to produce a functional
$\delta$-function with the effect of imposing the constraint
\begin{equation}
  \label{eq:ffdual}
  F_{\mu \nu}\, \widetilde{F}^{\mu \nu} =0 \nn.
\end{equation}
With such a constraint imposed on the gauge field it would of course
be no wonder if one gets no anomalies. In fact it would mean that 
the anomaly
\begin{equation}
  \label{eq:anomalyff}
\partial_{\mu}\, j^\mu \propto F_{\mu \nu}\, \widetilde{F}^{\mu \nu}
\end{equation}
had been constraint to be zero. Such a constraint will lead to
interactions between photons which are of course in the next
approximation in $m^2$ understandable as due to exchange of the
$b$-particles, $\ie$, $\gamma \gamma \to \gamma\gamma$. However,
notice that diagrams, like Fig.~\ref{fig:photonb}, have the $b$ field
propagator which has the contribution of a factor $m^{-2}$. This
propagator contribution is tremendous compared to $p^{-2}$ since $p$
is in the range (orders of magnitude) above the $U(1)_X$-photon mass
scale, $m$, $\ie$, $\gamma\gamma$-scatterings have extremely strong
interactions. Therefore, they may be able to provide the constraint
forces which uphold the constraint Eq.~(\ref{eq:ffdual}) (or
Eq.~(\ref{eq:ffmatrix})).

\subsection{Equations for the four dimensional effective theory}
\label{sec:eqsfoureffec}

We have an interacting $U(1)_X$-photon theory with the
Lagrangian
\begin{equation}
  \label{eq:blag}
  \mathscr{L} = - \frac{1}{4}\, F_{\mu\nu} \,F^{\mu\nu}
+ \frac{b}{16 \pi^2}\,  F_{\mu\nu} \,\widetilde{F}^{\mu\nu} \nn,
\end{equation}
which in addition has the term (\ref{eq:YMKin}) when we do not consider
$m$ so small that we can ignore the term in Eq.~(\ref{eq:YMKin}).

The equations of motion become, in addition to Eq.~(\ref{eq:ffdual})
derived by varying $b$,
\begin{equation}
  \label{eq:eomb}
  \partial_\mu F^{\mu\nu} = \frac{1}{4 \pi^2} \, 
\left(\partial_\mu b\right)\, 
\widetilde{F}^{\mu\nu} \nn.
\end{equation}

Including charged matter and noticing that the no-matter terms in the
Euler-Lagrangian equation can be written in a form more familiar, we
obtain
\begin{eqnarray}
 \label{eq:redeleq}
  \partial_\mu F^{\mathrm{Red}~\mu\nu} &=& \partial_\mu \left(
F^{\mu\nu} - \frac{b}{4 \pi^2}\, \widetilde{F}^{\mu\nu} \right) \\
  \label{eq:redeleq1}
&\equiv& J^\nu \nn,
\end{eqnarray}
where $J^\nu$ is the ``matter current'', and we may define the short
hand notation
\begin{equation}
  \label{eq:shandred}
 F^{\mathrm{Red}~\mu\nu} = F^{\mu\nu} - \frac{b}{4 \pi^2}\, 
\widetilde{F}^{\mu\nu}\nn. 
\end{equation}

{}For future discussions it is convenient when 
we express the equation of motion (Eq.~(\ref{eq:eomb})) with 
the electric field, $\vec{E}$, and magnetic field, $\vec{B}$,
\begin{equation}
  \label{eq:divERad}
  \mathrm{div} \vec{E} = \frac{1}{4 \pi^2}\, 
\mathrm{div}\left( b\, \vec{B}\right) \nn,
\end{equation}
hereby we have applied $\mathrm{div} \vec{E}^\mathrm{Red} =0$.

A mathematical point worth noticing in connection with the somewhat
unusual electrodynamics, which we discuss here, is that the condition
(\ref{eq:ffdual}) can be shown by trivial algebra to imply
that even
\begin{equation}
  \label{eq:ffmatrix}
   F_{\mu \nu}\, \widetilde{F}^{\nu\rho} =0 \nn,
\end{equation}
for all combinations of the indices of $\mu$ and $\rho$.

\subsection{Order of magnitude possibilities}
\label{sec:order}

Above, we have reviewed an anomalous horizontal $U(1)_X$ model
using Fayet-Iliopoulos D-term to cause a spontaneously break down of
$U(1)_X$-charge. This effect would arise if the dilaton field $\phi$
were non-zero which is, however, not achievable due to supersymmetry
being driven to be an exact symmetry. However, if we have a non-zero 
dilaton field, we also have world sheet instanton
effects breaking the global charge conservation using Green-Schwarz
anomaly cancellations. Thus, this suggests to use this instanton
effect in stead of a Fayet-Iliopoulos D-term. Could this effect be
adjusted for a phenomenologically reasonable global charge breaking of
order of the Cabbibo angle $\epsilon$?

Baring a totally mysterious cancellation of various contributions from
different world sheet instantons the order of magnitude of the
$U(1)_X$ breaking by world sheet instantons is estimated as the
exponent of the supersymmetry associated term to the field $b$, which
makes up a complex field. The possibility of surprising cancellations
would have \mbox{{\it a priori}} to be ignored~\cite{globalgs3} if it
were not for the findings that this indeed easily can
occur~\cite{Witten2}. Let us divide the discussion into the following
possibilities:
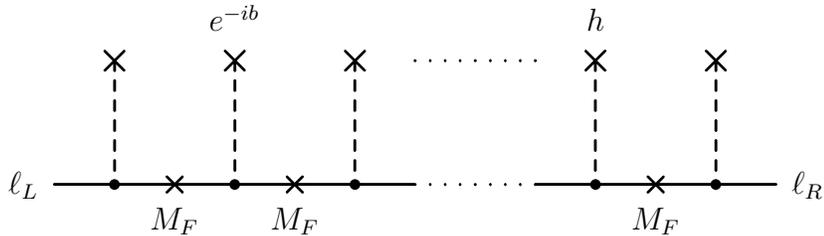
\begin{figure}[t!]
\unitlength=0.3mm  
\vspace{10mm}
\begin{center}
\begin{fmffile}{FNdiagram}
\begin{fmfgraph*}(320,55)
\fmfstraight
 \fmfleft{i2,i1}
 \fmfright{o2,o1}
\fmfv{label=$e^{-i b}$,l.a=90,l.d=4mm}{v3}
\fmfv{label=$h$,l.a=90,l.d=4mm}{v9}
\fmfv{label=$\ell_R$,l.a=0}{o2}
\fmfv{label=$\ell_L$,l.a=180}{i2}
\fmfv{label=$M_{F}$,l.a=-90,l.d=3mm}{v11}
\fmfv{label=$M_{F}$,l.a=-90,l.d=3mm}{v13}
\fmfv{label=$M_{F}$,l.a=-90,l.d=3mm}{y2}
\fmf{phantom}{i1,v1,v2,v3,v4,v5,v6,v7,v8,v9,y1,y3,o1}
\fmffreeze
\fmf{plain}{v10,i2}
\fmf{plain}{v10,v11}
\fmf{plain}{v11,v12}
\fmf{plain}{v12,v13}
\fmf{plain}{v13,v14}
\fmf{plain}{v14,v15}
\fmf{dots}{v15,v16}
\fmf{dots}{v16,v17}
\fmf{dots}{v6,v7}
\fmf{dots}{v7,v8} 
\fmf{plain}{v17,v18}
\fmf{plain}{v18,y2}
\fmf{plain}{y2,y4}
\fmf{plain}{y4,o2}
\fmf{dashes,tension=0}{v1,v10}
\fmf{dashes,tension=0}{v3,v12}
\fmf{dashes,tension=0}{v5,v14}
\fmf{phantom,tension=0}{v7,v16}
\fmf{dashes,tension=0}{v9,v18}
\fmf{dashes,tension=0}{y3,y4}
\fmfv{decor.shape=circle,decor.filled=full,decor.size=3thin}{v10,v12,v14,v18,y4}
\fmfv{decor.shape=cross,decor.filled=full,decor.size=10thin}{v1,v3,v5,v9,y3}
\fmfv{decor.shape=cross,decor.filled=full,decor.size=8thin}{v11,v13,y2}
\end{fmfgraph*}
\end{fmffile}
\vspace{8mm}
\end{center}
\caption{\protect Feynman diagram for fermion mass term. The dashed lines with 
  crosses symbolise of different Higgs fields ($h$) or the world sheet
  instanton ($e^{-i b}$). $M_F$ is denoted as the fundamental scale.}
\vspace{3mm}
\label{fig:FNdag}
\end{figure}

\begin{description}
\item[{\it (1)}] All the quantities, including $g_s=\phi^2$, are very
  strictly of order unity and the breaking of the charge conservation
  is also of order unity in spite of the fact that it is exponentially
  suppressed -- as an instanton tunnelling effect. In this philosophy the
  charge conservation is strongly broken: Let us then imagine that the
  mass or the effective Yukawa coupling for a quark or a charged
  lepton is obtained via a chain diagram (Fig.~\ref{fig:FNdag}) in
  which a series of fundamental scale vector coupled fermion
  propagators are linked by Higgs fields or world sheet instanton
  caused transition symbols. If the strength of the charge violating
  world sheet instantons is of just the same order of magnitude as the
  (typical) fundamental scale fermion masses, then there will be no
  suppression, and the $U(1)_X$-charge considered will be of no help
  in explaining the suppression of some effective Yukawa couplings (at
  experimental scales) compared to others.  However, taking
  ``everything'' especially the compactifying space dimensions to be
  very close to unity in ``fundamental'' units, such that even
  exponents are accurately of order unity is presumably not likely
  to be true.
\item[{\it (2)}] The other possibility is that there are some
  quantities which cannot be considered order unity in the very strong
  way discussed under point {\it (1)}. The compactifying dimensions
  turn out to be the quantities of importance for the strength of the
  $U(1)_X$-charge violation. Let it be clear that there are really two
  scales of breaking of the $U(1)_X$ symmetry to be discussed:
\begin{description}
\item[{\it (a)}] There is the $U(1)_X$-photon mass scale, $\ie$, the
  mass $m$ of Eq.~(\ref{eq:YMKin}).  
\item[{\it (b)}] The mass scale $M_V$ that appear as the mass obtained
  for the fermions which are mass protected {\em only} by the $U(1)_X$
  and get their mass via the world sheet instanton caused term
  (\ref{eq:instop}) in the case of $Q={\overline\psi}\, \psi$.  It is
  this mass scale $M_V$ which divided by the ``fundamental'' fermion
  masses $M_F$ gives the suppression factor $\epsilon=M_V/M_F$ which is
  used as a factorisation parameter for the fermions mass matrices.
\end{description}
\end{description}

We discuss in the following the variation of the scales of breaking 
{\it (a)} and {\it (b)} above: The mass square factor in
Eq.~(\ref{eq:YMKin}) goes back to the term (\ref{eq:kinKR}) in as far
as the $b$ field is a coefficient on a term in $B$ which in turn has
its derivatives go into $H$ in Eq.~(\ref{eq:kinKR}). It is remembered
that the function multiplying $b$ to get the $B_{45}$ contribution
must be normalised so that a shift in $b$ by $2\pi$,
\begin{equation}
  \label{eq:bshift}
  b \to b + 2\pi \nn,
\end{equation}
corresponds to adding to $\int_{S^2} B$~the shifting by a single
monopole flux field through the cycle $S^2$ shifted by $\Lambda=2\pi$.
If all the couplings are taken to be of order unity, one finds that
scaling the dimensions of the $2$-cycle as the typical compactifying
space dimension, $R$, squared $R^2$. The area of the $2$-cycle is
proportional to $R^2$. This means that that $m^2\propto R^{-2}$. Thus
we see that the mass scale of the $U(1)_X$-photon goes as $R^{-1}$
where $R$ is really the length scale of the two-cycle.

The tunnelling suppression amplitude of the world sheet 
instanton~\cite{Witten1,Witten2} is
\begin{equation}
  \label{eq:insttunn}
  \exp\left( - \frac{A}{2\pi \alpha^\prime} 
+ i \int B\right) \frac{\mathrm{Pfaff}^\prime \mathscr{D}_F}
{\sqrt{\mathrm{det}^\prime \mathscr{D}_B}} \nn,
\end{equation}
where $\alpha^\prime$ is the Regge slope, $\mathrm{Pfaff}$ is the
Pfaffian, and $\mathscr{D}_F$ and $\mathscr{D}_B$ are kinetic
operators for the bosonic and fermionic fluctuations, respectively.
The ``$\hspace{1mm}{}^\prime\hspace{1mm}$'' on the 
Pfaffian and determinant denotes that the
zero modes are to be omitted. Moreover, $A$ is the area, which of
course $A\propto R^2$ again, with $R$ being the $R$ relevant for
the two-cycle. 

Introducing a fundamental mass scale, $M_F$, we thus
have the scale at which the $U(1)_X$-charge conservation
violates
\begin{equation}
  M_V \approx M_F \, \exp\left( - R^2\ M_F^2 \right) \nn.
\end{equation}
This crude estimate assumed $g_s$ to be of order one.

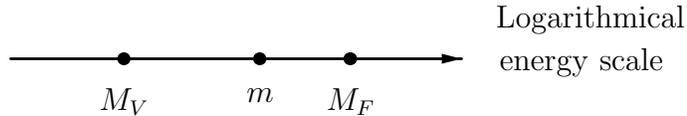
\begin{figure}[t!]
\begin{center}
\unitlength=1mm
\begin{fmffile}{Energylev}
\begin{fmfgraph*}(60,30)
\fmfleft{i}
\fmfright{o}
\fmffreeze
\fmf{plain}{i,i1,i2,i3,i4,i5,i6,i7,i8,i9,i10,i11,i12,i13,i14,i15,%
i16,i17,i18,i19,o}
\fmfdot{i5,i11,i15}
\fmf{plain_arrow, tension=0.1}{i19,o}
\fmfv{label=$M_F$,l.a=-90,l.d=4mm}{i15}
\fmfv{label=$\mathrm{Logarithmical}$,l.a=25,l.d=8mm}{i19}
\fmfv{label=$\mathrm{energy~scale}$,l.a=-5,l.d=5mm}{o}
\fmfv{label=$m$,l.a=-90,l.d=4mm}{i11}
\fmfv{label=$M_V$,l.a=-90,l.d=4mm}{i5}
\end{fmfgraph*}
\end{fmffile}
\end{center}
\caption{\protect The different energy scale. $M_F$ is 
the fundamental scale. $m$ 
is $U(1)_X$-photon mass. $M_V$ is $U(1)_X$ breaking 
scale. Suggestive rations: $M_F/m\approx1.5$,  $M_F/M_V\approx3$.}
\vspace{3mm}
\label{fig:energylev}
\end{figure}

Now, we may go into crude phenomenology, still taking $g_s$ of order
unity. In theories with compactified extra dimensions it is quite
natural to take this as being the reason for the fine structure
constants being weak, compared to the ``self-dual'' strength
(defined as the value of fine structure constant which makes it equal
to the corresponding monopole coupling constant, associated by the
Dirac relation). The self-dual strength,
${\widetilde\alpha}_{U(1)_X}$, is approximately $1/2$ since
$\alpha_e\,\alpha_g=1/4$ for a formal monopole $\alpha_g$. From this
point of view we can claim that a typical, say GUT, coupling of order
$\alpha\approx 1/25$ is weaker by a factor $\approx12$ than the
Abelian self-dual. Though this is exaggerating and should be corrected
at least by the factor $3/5$. Roughly taking anything of this order,
we would now expect that $R$ measured in ``fundamental units'' would
be of the order $R\approx \sqrt[6]{12}\simeq1.5$ (see
Fig.~\ref{fig:energylev}). This would mean $m\sim M_F/1.5$ and
$M^2_V\sim M_F^2 \exp(-1.5^2)$, $\ie$, $M_V \sim M_F/3$, a very useful
suppression factor indeed. This is namely a typical order of magnitude for an
$\epsilon$ with which one fits the fermion mass
spectra~\cite{globalgs2,globalgs3}.

In this way it looks very promising to get very naturally a good scale
for violation strength for phenomenological fitting. We can say that,
strictly speaking, a scale $R$ of the compactifying dimensions just
$1.5$ times bigger than the fundamental length scale $M_F^{-1}$ is so
close to being of order unity, that there is hardly any call for any
special explanation for this ``deviation'' from it being the
fundamental size. That we can notice some small numbers in both the
fine structure constants and in the suppressed violation of the
$U(1)_X$-charge is due to respectively the $6$ dimensional
compactifying space and the exponentiation because of the world sheet
instanton effect needed.

The energy scale gap in which we have the funny electrodynamics with
the constraint $F_{\mu\nu}\, \widetilde{F}^{\mu\nu}=0$ is in the just
sketched scenario (with $g_s \sim 1$) reduced to a scale factor in the
$1.5$ region. That is such a very small range that one would hardly be
able to claim that anything strange will happen.

\subsection{Wormhole discussion I}
\label{sec:wormholeI}

Independent of whether world sheet instantons do or do not break
global charge, there is another mechanism threatening the global charge
conservation: the effect of wormholes.

We shall in fact show below in Subsec.~\ref{sec:wormholeII} that our
estimation of the effects of gravitational wormholes present in the
vacuum (or similar space-time foam effects) {\em will give rise to a
  significant violation of the global charge -- in the four
  dimensional theory -- with the non-trivial Green-Schwarz anomaly
  cancellation, although it has resulted from a gauge charge} and thus
at first might have been suspected of not being violated by space-time
foam effects.

Over most of the energy scale (logarithmical counted, see
Fig.~\ref{fig:energylev}) between the fundamental scale and the
$U(1)_X$ violation scale $M_V$, we have a globally conserved charge
for $U(1)_X$ while the gauge particle ``already'' had its mass at
$m\sim R^{-1}$.

Now, however (see Subsec.~\ref{sec:wormholeII}), according to
arguments in Ref.~\cite{Gilbert} charges which are not gauge protected
may disappear into wormholes or baby universes. Therefore, one may
speculate whether such a charge conservation which is unprotected by
gauge fields will not get its conservation spoiled by the Wheeler
space-time foam. Actually such effects of breaking the global charge
are expected at energy scales smaller than $m$. But we will see in
Subsec.~\ref{sec:wormholeII} that wormholes at even higher energy
scales than $m$ shall provide such breaking when the monopoles are
used. The wormholes of low energy (smaller than $m$), $\ie$, of large
(length) are suppressed exponentially with some exponent related to
$m/M_F$. In fact we would (naively) estimate that for a space-time
foam ingredient such as a baby universe of size $m^{-1}$ (in length)
we would have an action of order $M_F^2/m^2$, and a suppression factor
of $\exp(-M_F^2/m^2)\approx\exp(-R^2M_F^2)$. It happens under the
assumption $g_s \approx 1$ to be just the same order of magnitude --
in the exponent suppression factor -- as the one present in the world
sheet instanton suppression factor. For this coincidence to occur it
was quite crucial that the estimate just used for the baby universe
action was of the form $\int_C R~\sqrt{g}~d^4 x$ ($C$ is baby universe
tube) as being obtained by use of Einstein-Hilbert action and not
$\int_C \sqrt{g}~d^4 x$ which would have been the case if the
cosmological term in the action were significant here. That is to say,
we used that the cosmological constant is zero, but at these scales
one is closer to the (running) cosmological constant which is relevant
for short distance and which may have another value than the long
distance one which is practically zero. In any case even if the
cosmological constant were relevant, it would suppress the baby
universes at the scale $m$ even more.

The space-time foam non-conservation is expected to be at most what the 
world sheet instanton effect would have been if the 
$g_s$ were of order unity, contrary to true 
expectation. Now, however, since 
we truly do not expect the world sheet instantons to provide breaking 
then the wormhole effects could easily take over as the dominant effect.
Rather we should say that we do expect an appreciable wormhole breaking
(see the argument below).

It could be that the world sheet breaking effect could actually
dominate even in the case of $g_s \approx1$, but the result would in
this case not be so different with respect to the order of magnitude
of the (exponent of the) effect, and thus the conclusion would be the
same. However, in the case that there is no significant world sheet
instanton effect, the wormholes can very well work and completely and
dominantly break the charge conservation.

\subsection{Argument for wormhole effect violation of the global part of the 
  $U(1)_X$-charge -- Wormhole discussion II}
\label{sec:wormholeII}

In this subsection we will prove for that the $U(1)_X$-charge
is violated as a global charge due to the wormhole effects.

In the foregoing subsection we saw that one could give immediate
arguments both in favour of and against the global $U(1)_X$-charge being
violated by the wormhole or space-time foam effects. In this subsection we
want to deliver an argument which shows that indeed {\em the global
  $U(1)_X$-charge is broken due to the wormholes}. We should, however,
stress that this breaking is exponentially suppressed but {\em only
  with the suppression corresponding to the scale of the Higgsing
  $m$}, which is essentially the compactification scale. In the
realistic models this is a rather high energy scale and the breaking
from wormholes is thus expected to be quite large. Compared to what
one gets from world sheet instantons, which do not work after all in
the case of $g_s$ being of order one, this could be much bigger than
the latter.

It is most easy to organise the non-conservation of the Coulomb field
for the $U(1)_X$-charge by use of virtual wormhole entrances with
magnetic fluxes radiating out. These form effective virtual magnetic
monopoles in the vacuum. The reason that it is profitable with
monopoles violating the charge conservation for the $U(1)_X$-charge
is that we indeed can derive some formulas for the
variation/development of the $U(1)_X$-electric-charge relating it to
the variation of $b$ on the sites of the monopoles (see
Eq.~(\ref{eq:divERad})),
\begin{equation}
  \label{eq:delzeroE}
\partial^0 \mathrm{div}{} \vec{E} =
\frac{1}{4 \pi^2} \, \partial^0 \mathrm{div}{} 
\left( b \vec{B} \right)\nn.
\end{equation}

Since $\mathrm{div} {}\vec{E}$ is the charge density this formula
tells us that, for instance, when a monopole is present, the variation
rate of the charge $\partial^0 \mathrm{div}{} \vec{E}$ contains a term
$(\partial^0 b)\, \mathrm{div}{} \vec{B}$ on the site of a monopole
whenever $b$ varies, $\ie$, $\partial^0 b\not=0$. The monopoles to be
used here do not have to be genuine monopoles. They could be entrances
to wormholes with magnetic flux going through~\cite{Wheeler}. We would
not even have to use such genuinely existing monopolic wormhole
entrances. It is rather sufficient to consider entrances which are
virtually present in the vacuum. We shall imagine that there are many
such wormholes virtually present with magnetic flux and that the
entrances give rise to interactions with the various fields in the
theory. We assume that interactions can be described by effective
terms in the Lagrangian density. At first they are only at the places
where the entrances to the wormholes are. We shall, however, integrate
over all the possible positions or movements for the wormholes -- as
part of the functional integration of Feynman path integral. This has
the implication that one can achieve naturally that such a model of
wormholes can become effectively translational invariant. In fact one
has to integrate over the positions -- of the wormhole or baby
universe entrances -- with a translational invariant measure.  That
can actually be supported by a Heisenberg-inequality type argument
using that at least baby universes cannot transport energy and
momentum, because the information about these quantities is safely
stored in the gravitational field at long distances from a (supposed
to be) little baby universe. Even for wormholes it is reasonable to
integrate over all positions with a translational invariant measure.

Since it is clearly possible that any sort of particle could be
scattered into a wormhole, the effective Lagrangian density
contribution from the entrance to a wormhole can contain terms
annihilating or creating any combination of particles. Thus any
combination/product of fields is {\it a priori} possible and will come
with some coefficient in the effective wormhole and baby universe
Lagrangian. Now, however, there can be processes that cannot really
take place due to Coulomb fields left behind. By this we mean that if
we propose terms violating gauge symmetries for charges with light
gauge particles associated there remains information outside. In fact
there will be a Coulomb field left, carrying the information about the
charge of the particles which have gone into the wormhole. Even if the
particle goes deeply into the wormhole and maybe even out somewhere
else far away, there will remain electric flux lines exiting from the
entrance of the wormhole and even if no appropriate -- may be
different -- particle is pulled out the entrance the wormhole itself
will behave as a charged particle. In this way we can only have
effective Lagrangian density terms conserving the gauged charges
corresponding to gauge particles with Compton wave lengths which 
are long compared to the wormhole sizes.

Really what matters is, whether the gauge field around the wormhole
can keep the information about what went into it. In the case of the
conserved $U(1)_X$-charge the $\mathrm{div}{} \vec{E}$, that should
have ensured the stability of the Coulomb field, does not correspond to
a conserved current as in the usual electrodynamics. It is rather the
current corresponding to the $F^{\mathrm{Red}\, \mu\nu}$
(Eq.~(\ref{eq:shandred})) that is conserved. In fact we have just seen
that if the axion field $b$ varies on the sites of the virtual
monopoles we can/will have that $\mathrm{div}{} \vec{E}$ and thus the
charge varies.

In this way it should now be {\em allowed} to have Lagrangian density
terms due to the monopolic wormholes violating the $U(1)_X$. Once such
terms are allowed they are expected to be there and we will generate
masses for particles which are only mass-protected by $U(1)_X$ group.

Once we have the symmetry strongly broken at the Planck scale which is
now expected there will no longer be a sign of the conservation and
thus also no problem with the anomalies. The symmetry seems to be
dynamically broken -- not only spontaneously -- because the effective
Lagrangians representing the wormhole and other space-time foam
effects really have to be interpreted as dynamical breaking. We must
also expect that it is rather impossible to keep the $U(1)_X$-photon
mass $m$ to be light compared to $M_V$ under such conditions.

We conclude that seriously taking wormholes into account in this
way, it results that the Green-Schwarz anomaly cancellation scheme 
{\em does not work } in $3+1$ dimensional limit.

What happens is that the very strong constraint ensuring forces due to
very large $b$ propagators lead to the possibility of 
Coulomb fields around a wormhole entrance modified with time. This
modification possibility in turn allows the effective Lagrangian
density corresponding to the absorption of the 
$U(1)_X$-charges into wormholes.

\section{Conclusions}
\label{sec:conclusions}

Anomaly cancellation by the Green-Schwarz mechanism in the case of a
certain $3+1$ dimension limit of a higher dimensioned string theory is
questioned: We consider the gauge symmetry (needed to the $U(1)_X$
subgroup) that allegedly results from breaking a larger string theory
gauge group using a field $b$ derived from Kalb-Ramond $B_{M N}$ that
takes on a non-vanishing vacuum expectation value and thereby higgses
the gauge field $A_{\mu}$. This is manifested phenomenologically as an
approximately conserved current without having the usual triangle
summation anomaly requirement for avoiding gauge and mixed anomalies.
This is referred to as Green-Schwarz anomaly cancellation because the
special way of having anomaly cancellations for string theory mass
states is inherited to the $3+1$ dimensional limit.

If we have supersymmetry and such a charge with Green-Schwarz anomaly
cancellation (effectively in the $3+1$ dimensions), then according to
the calculations of Ref.~\cite{GS4} we have a Fayet-Iliopoulos D-term
(\ref{eq:anomalousFI}) which drives the dilaton field that is
essentially in correspondence with the effective string coupling
constant $g_s=\phi^2$ to be zero. That means that the theory becomes
{\em free} and thus rather unrealistic. This is a severe
trouble in itself for the models with a non-trivial anomaly cancellation
mechanism.

However, contrary to this argument, if we should assume that in
some mysterious way it were possible to get a string theory after all
with supersymmetry surviving and having a significant D-term and string
coupling in spite of such Green-Schwarz cancellation, then the world
sheet instantons would prevent the breaking strength of the surviving
global charge conservation from going to zero.

It is a major point of the present article that it is actually not
achievable for the world sheet instantons do violate the
$U(1)_X$-charge conservation. In this way we could avoid the mystery
of having a gauge theory breaking from a spontaneous symmetry
breaking, which does not break the current conservation. Provided that
the world sheet instanton effects do not mysteriously cancel (which is
though less safe to assume than naively expected), this mystery would
disappear.

The authors working with the Green-Schwarz anomaly cancelling
charges (usually) have in mind supersymmetric models. If the world
sheet instanton effects would appear even in supersymmetric theories,
the Green-Schwarz anomaly cancellation would become less suspicious of
being pretended to behave strangely from the general point of view.

However, due to the Fayet-Iliopoulos D-term expected when we have such
charges, the string coupling constant $g_s$ may be driven to zero and
the whole effect of world sheet instantons would disappear. This may
though be not realistic because of bringing the whole string
theory appears to be free.

If this kind of effect could work causing a breaking of the global
charge conservation and thus avoiding some of the strangeness one may
still seek (using the world sheet instantons) to declare a
$U(1)_X$-charge needing the Green-Schwarz anomaly cancellation to have
become less suspicious of being pretended to behave strangely from
general point of view. However, one may still declare the way to be
suspicious, in which the anomaly troubles are avoided in the energy
range above the mass scale $m$ of the gauge particle: The gauge fields
are constrained to never have the configuration leading to anomalies!

If one gets a breaking of charge $U(1)_X$ due to the wormholes (or
world sheet instantons), one could imagine as an
interesting possibility to {\em use this breaking} instead of some
spontaneous breaking induced from $\eg$ the Fayet-Iliopoulos D-term
(as \cite{globalgs3} proposes) to provide the soft breaking which is
needed to use the charge $U(1)_X$ as a mass protecting charge to
implement the large ratios of quark and lepton masses.

We further expressed our worry and suspicion whether such an
electrodynamics which is constrained by
$F_{\mu\nu}\widetilde{F}^{\mu\nu}=0$ can really be considered as
realistic on general physics grounds or whether it represents an
unrealisable speculation concerning the relative orders of magnitude.

However, we thought there are reasons to believe that the wormholes
would give violation of the $U(1)_X$ even above its Higgsing scale.
That could mean that wormholes break the $U(1)_X$ symmetry completely
under the use of $F_{\mu\nu}\widetilde{F}^{\mu\nu}=0$. The point is
that a very strong coupling of the $b$ field causes that Coulomb field
around the wormholes endings are not stable.

\ack{We wish to thank S.~Lola for useful discussions, and Y.T. would
  like to thank F.~Bigazzi and K.~S.~Narain for useful conversations
  on the heterotic strings.  We also wish to thank M.~de Riese for
  careful reading of this manuscript. H.B.N. thanks the Abdus Salam
  International Centre for Theoretical Physics, and Y.T. thanks the
  Niels Bohr Institute for hospitality during the preparation,
  respectively.}
%
%

\end{document}